\documentstyle[12pt]{article}
\begin{document}

\noindent {\small CITUSC/00-019\hfill \hfill hep-th/0004090 \newline
}

{\small \hfill }

{\vskip0.5cm}

\begin{center}
{\Large {\bf 2T Physics Formulation of Superconformal Dynamics}}

{\Large {\bf Relating to Twistors and Supertwistors}}{\footnote{%
This research was partially supported by the US. Department of Energy under
grant number DE-FG03-84ER40168.}}{\Large {\bf \ }}

\bigskip

{\vskip0.5cm}

{\bf Itzhak Bars}

{\vskip0.5cm}

{\bf CIT-USC Center for Theoretical Physics}

{\bf and}

{\bf Department of Physics and Astronomy}

{\bf University of Southern California}

{\bf \ Los Angeles, CA 90089-2535, USA}

{\vskip0.5cm}

{\bf Abstract}
\end{center}

The conformal symmetry SO$(d,2)$ of the massless particle in d dimensions,
or superconformal symmetry OSp$\left( N|4\right) $, SU$(2,2|N)$, OSp$(8|N)$
of the superparticle in $d=3,4,6$ dimensions respectively, had been
previously understood as the global Lorentz symmetry and supersymmetries of
2T physics in $d+2$ dimensions. By utilizing the gauge symmetries of 2T
physics, it is shown that the dynamics can be cast in terms of superspace
coordinates, momenta and theta variables or in terms of supertwistor
variables \`{a} la Penrose and Ferber. In 2T physics these can be gauge
transformed to each other. In the supertwistor version the quantization of
the model amounts to the well known oscillator formalism for non-compact
supergroups.

\section{2T Physics}

Two-time (2T) physics has by now been shown to provide a reformulation of
all possible one-time (1T) particle dynamics, including interactions with
Yang-Mills, gravitational and other fields \cite{conf}-\cite{2Tfield}. 2T
physics has mainly been developed in the context of particles, but some
advances have also been made with strings and p-branes \cite{string2t}, and
some insights for M-theory have already emerged \cite{liftM}\cite{toyM}. In
the case of particles, there exists a general worldline formulation with
background fields \cite{emgrav}, as well as a field theory formulation \cite
{2Tfield}, both described in terms of fields that depend on $d+2$
coordinates $X^{M}$. The 2T point of view has been useful in bringing new
insights into 1T physics, first by revealing previously unnoticed hidden
symmetries in 1T dynamical systems, and second by providing a unification of
classes of 1T dynamics that are different in 1T physics, but are regarded as
being gauge equivalent to each other under the local gauge symmetries of a\
unique 2T system.

It is important to realize that the Standard Model of particle physics and
gravity (i.e. all the physics we can verify) can be rewritten as a 2T field
theory, modulo a reformulation of the mass term in the Higgs potential. The
2T approach is not a naive extension of the number of timelike dimensions,
as this would be disastrous for unitarity and causality, but it does imply
the existence of a more subtle higher structure of dimensional unification,
including an extra timelike and an extra spacelike dimension, {\it plus
gauge symmetries} that tend to reduce the number of effective dimensions
(worldline Sp$\left( 2,R\right) $ or OSp$\left( n|2\right) ,$ and local
spacetime generalized bosonic/fermionic kappa symmetries). Thus, for a fixed
set of 2T background fields (e.g. flat background) one can find many related
interacting 1T systems by gauge fixing the underlying gauge symmetry. The
non-trivial aspects of 2T physics is due to the fact that $d$ dimensions
(1T) can be embedded in many ways inside $d+2$ dimensions (2T) (also such
concepts extended to superspace). The 1T observers see the same 2T system as
different dynamics from the point of view of the chosen $d$ dimensions, but
all 1T observers are gauge equivalent from the point of view of the 2T
observer. This gauge equivalence is translated to a kind of ``duality''
among the 1T dynamical systems that are derived from a given 2T system.

The origins of the field theory formalism goes back to Dirac's work in 1936
on conformal symmetry, but this approach historically was used exclusively
as a reformulation of conformal symmetry in field theory \cite{Dirac}-\cite
{vasilev}. An early approach on the worldline is also aimed to conformal
symmetry \cite{marnelius}. The unification of different dynamics in the form
of 2T physics was not understood until the work of \cite{conf}, which
reached the Sp$\left( 2,R\right) $ gauge theory formulation on the worldline
by following a very different path and motivation (coming from 2T signals
and dualities in M-theory, F-theory, S-theory, and influenced directly by
the formalism in \cite{kounnas}). Also, in the field theory formalism, it
was realized only recently \cite{2Tfield} that the 2T field equations are a
reflection of the Sp$\left( 2,R\right) $ gauge symmetries that underlie 2T
physics on the worldline, and that the gauge fixing procedure which produce
the various 1T dynamics in the worldline formulation can be carried out in
the field theory formalism as well. Although the 2T unification of 1T
systems can be examined in either the worldline or field theory forms, the
worldline approach provides a better understanding of the underlying gauge
symmetries at this stage, while the field theory formulation provides a
familiar approach for interactions among fields.

In this paper we will further extend the gauge symmetries of 2T physics on
the worldline by further developing some concepts already introduced in \cite
{super2t} that were necessary for the 2T formulation of supersymmetry. The
gauge symmetry is associated with the spacetime SO$\left( d,2\right) $, some
internal symmetry and an extended concept of bosonic/fermionic kappa
supersymmetry. We will use the Sp$\left( 2,R\right) $ gauge symmetry
together with the extra gauge symmetry to show that particle dynamics in one
of the Sp$\left( 2,R\right) $ 1T gauges (the SO$\left( d-1,1\right) $
relativistic particle gauge\footnote{%
The same idea could be applied in other Sp$\left( 2,R\right) $ 1T gauges,
but we will refrain from discussing other Sp$\left( 2,R\right) $ gauges in
this paper.}) can be cast either in terms of particle coordinates and
momenta $\left( x^{\mu },p^{\mu }\right) $ or in terms of twistor variables
\`{a} la Penrose \cite{penrose} or supertwistors \`{a} la Ferber \cite
{farber}. This result will be extended to the conformal superparticle cases
for $d=3,4,6$ with $N$ supersymmetries using the supergroups OSp$\left(
N|4\right) ,$ SU$\left( 2,2|N\right) ,$ OSp$\left( 8^{\ast }|2N\right) $
respectively. It will be shown that the (super)particle description $\left(
x,p,\theta \right) $ is {\it gauge equivalent} to the (super)twistor
description when the SO$\left( d,2\right) $ and Sp$\left( 2,R\right) $ gauge
symmetries are utilized. Furthermore, it will be shown that the quantization
of the model can be carried out in terms of oscillators associated with the
(super)twistors, and the unitarity established. The main motivation for
developing this formalism came from trying to understand the quantum theory
of a toy M-model introduced in \cite{liftM}. The methods presented here will
be applied in \cite{toyM} to the case of \ the toy M-model based on the
supergroup OSp$\left( 1|64\right) .$

\section{1T superparticle}

Since our ultimate goal includes spacetime supersymmetry we will directly
discuss the massless superparticle; the reader can delete the fermions to
specialize to the purely bosonic case. The action for the superparticle in $%
d $ dimensions, with $N$ supersymmetries, is traditionally described by the
following well known Lagrangian written in terms of position $x^{\mu }\left(
\tau \right) $, momentum $p^{\mu }\left( \tau \right) $ and $N$ spacetime
spinors $\theta _{\alpha }^{a}\left( \tau \right) $ ($\alpha $ denotes the
spinor and $a$=1,2,$\cdots ,N$) 
\begin{equation}
\pounds =\frac{1}{2A^{22}}\left( \dot{x}^{\mu }+\tilde{\theta}_{a}\gamma
^{\mu }\partial _{\tau }\theta ^{a}\right) ^{2}=\dot{x}\cdot p-\frac{1}{2}%
A^{22}p^{2}+\tilde{\theta}_{a}\gamma\cdot {p}\partial _{\tau }\theta ^{a}.
\label{L1}
\end{equation}
We used the symbol $A^{22}\left( \tau \right) $ for the einbein because of
its relation to the Sp$\left( 2,R\right) $ gauge fields $A^{ij}\left( \tau
\right) $ that we will see below. The general variation of the fields has
the form (up to total derivatives) 
\begin{eqnarray}
\delta L &=&\partial _{\tau }\left( \delta x-\delta \theta _{a}\gamma \theta
^{a}\right) \cdot p+2\delta \theta _{a}\left( p\cdot \gamma \right) \dot{%
\theta}^{a} \\
&&+\delta p\cdot \left( \dot{x}^{\mu }+\tilde{\theta}_{a}\gamma ^{\mu
}\partial _{\tau }\theta ^{a}-A^{22}p\right) -\delta A^{22}\frac{p^{2}}{2}
\end{eqnarray}
For general $d$ and $N,$ using the appropriate spinor, this Lagrangian is
symmetric ($\delta L=0$ up to total derivatives$)$ under local $\tau $%
-reparametrization, local kappa supersymmetry and global super Poincar\'{e}
symmetry. In particular the global supersymmetry with parameters $%
\varepsilon _{\alpha }^{a}$ is given by 
\begin{equation}
\delta _{\varepsilon }x^{\mu }=-\delta _{\varepsilon }\theta \gamma ^{\mu
}\theta ,\quad \delta _{\varepsilon }\theta =\varepsilon ,\quad \delta
_{\varepsilon }p^{\mu }=0,\quad \delta _{\varepsilon }A^{22}=0,
\end{equation}
and the local kappa transformation with parameters $\kappa _{\alpha
}^{a}\left( \tau \right) $ is 
\begin{equation}
\delta _{\kappa }x^{\mu }=\delta _{\kappa }\theta \gamma ^{\mu }\theta
,\quad \delta _{\kappa }\tilde{\theta}_{a}=\tilde{\kappa}_{a}\gamma \cdot
p,\quad \delta _{\kappa }p^{\mu }=0,\quad \delta _{\kappa }A^{22}=4\tilde{%
\kappa}\dot{\theta}.
\end{equation}
For the special dimensions $d=3,4,6$ this Lagrangian is also invariant under
dilations, conformal transformations and special superconformal
transformations, such that the full global symmetry is given by the
supergroups 
\begin{equation}
G=OSp\left( N|4\right) ,\,SU\left( 2,2|N\right) ,\,OSp\left( 8^{\ast
}|N\right)  \label{G}
\end{equation}
respectively, as shown in \cite{schwarz} for $d=3,4$ and in \cite{super2t}
for $d=6.$ The conformal subgroups in these dimensions are SO$\left(
3,2\right) =Sp\left( 4\right) ,$ SO$\left( 4,2\right) =SU\left( 2,2\right) ,$
and SO$\left( 6,2\right) =Spin\left( 8^{\ast }\right) $ respectively. For
the purely bosonic case this Lagrangian has global conformal symmetry SO$%
\left( d,2\right) $ for any $d.$

\section{2T formulation}

The conformal symmetry SO$\left( d,2\right) $ is a giveaway for 2T physics.
Indeed all of the above cases correspond to a special Sp$\left( 2,R\right) $
gauge choice of a 2T formulation (the SO$\left( d-1,1\right) $ relativistic
particle gauge) in which the SO$\left( d,2\right) $ Lorentz symmetry in flat
backgrounds in $d+2$ dimensions gets interpreted as the conformal symmetry
in 1T (it acquires other interpretations in other Sp$\left( 2,R\right) $
gauges). The 2T reformulation of the superparticle in $d=3,4,6$ dimensions
requires $d+2$ coordinates $X^{M}\left( \tau \right) $ and momenta $%
P^{M}\left( \tau \right) ,$ and a supergroup element $g\left( \tau \right)
\in G$ that contains fermions $\Theta _{\tilde{\alpha}}^{a}\left( \tau
\right) $ where $\tilde{\alpha}$ denotes the spinor in $d+2$ dimensions.
This spinor has double the size of the spinor $\theta _{\alpha }^{a}\left(
\tau \right) $ in $d$ dimensions, which is of course necessary if the SO$%
\left( d,2\right) $ is to be realized linearly in the 2T formulation. Thus,
compared to the 1T formulation there are extra degrees of freedom in $%
X,P,\Theta $ and in the bosonic sectors in $g\left( \tau \right) .$ If the
covariant 2T formulation is to be equivalent to the 1T formulation there has
to be various gauge symmetries and extended kappa supersymmetries to cut
down the degrees of freedom to the correct set. Following \cite{super2t}
this is beautifully achieved as follows.

The 2T Lagrangian is 
\begin{equation}
L=\dot{X}_{1}\cdot X_{2}-\frac{1}{2}A^{ij}X_{i}\cdot X_{j}-\frac{1}{s}%
Str\left( \Gamma _{MN}\dot{g}g^{-1}\right) L^{MN},  \label{L}
\end{equation}
where $X_{i}^{M}=\left( X^{M},P^{M}\right) $ is the Sp$\left( 2,R\right) $
doublet, $A^{ij}$ is the Sp$\left( 2,R\right) $ gauge potential, $\Gamma
_{M} $ are gamma matrices and $\Gamma _{MN}=\frac{1}{2}\left[ \Gamma
_{M},\Gamma _{N}\right] $ are the SO$\left( d,2\right) $ generators in the
spinor representation of dimension $s,$ the Cartan connection $\dot{g}g^{-1}$
projected in the direction of the SO$\left( d,2\right) \in G$ is coupled to
the SO$\left( d,2\right) $ orbital angular momentum $L^{MN}=\varepsilon
^{ij}X_{i}^{M}X_{j}^{N}=X^{M}P^{N}-X^{N}P^{M}$ which is Sp$\left( 2,R\right) 
$ gauge invariant. Although our interest in this paper is on the supergroups 
$G$ listed in (\ref{G}) the discussion of local symmetries below applies
also to any group or supergroup that contains SO$\left( d,2\right) $ as a
subgroup. In fact, for the toy M-model in \cite{liftM}\cite{toyM} the case
of interest is $G=OSp\left( 1|64\right) .$ In particular, for the purely
bosonic particle one can simply take $G=SO\left( d,2\right) .$

The following is an improvement of the symmetry discussion given in \cite
{super2t}. From the extensive discussions in \cite{conf}-\cite{toyM} we
already know that the Lagrangian above has local symmetry under Sp$\left(
2,R\right) $. Beyond this, it obviously has global symmetry under $G$ for
the transformation of $g\left( \tau \right) $ by {\it right multiplication}
by a global group element $g_{R}\in G$ 
\begin{equation}
X_{i}^{M}\rightarrow X_{i}^{M},\quad A^{ij}\rightarrow A^{ij}{\quad }%
,g\rightarrow gg_{R}.  \label{global}
\end{equation}
Furthermore, it has local SO$\left( d,2\right) $ Lorentz symmetry with
parameters $\varepsilon ^{MN}\left( \tau \right) $ under {\it left
multiplication} of $g$ in the spinor representation and transformation of $%
X_{i}^{M}$ in the vector representation 
\begin{equation}
\delta X_{i}^{M}=\varepsilon ^{MN}X_{iN},\quad \delta g=\frac{1}{4}%
\varepsilon ^{MN}\left( \Gamma _{MN}\,g\right) ,\quad \delta A^{ij}=0.
\end{equation}
The time derivatives of $\dot{\varepsilon}^{MN}$ produced by the two kinetic
terms in (\ref{L}) cancel each other. Moreover, for the cases in which there
is another subgroup in $G$ (such as the SO$\left( N\right) ,$ SU$\left(
N\right) ,$ Sp$\left( N\right) $ in (\ref{G})) with generators $T_{A}$ that
satisfy Str$\left( T_{A}\Gamma _{MN}\right) =0,$ there is a local symmetry
with parameters $\varepsilon ^{A}\left( \tau \right) $ under {\it left
multiplication} of $g$ 
\begin{equation}
\delta g=\varepsilon ^{A}\left( T_{A}\,g\right) ,\quad \delta
X_{i}^{M}=0,\quad \delta A^{ij}=0.
\end{equation}
The time derivative $\dot{\varepsilon}^{A}$ as well other dependence on $%
\varepsilon ^{A}$ drops because Str$\left( T_{A}\Gamma _{MN}\right) =0$ and $%
\left[ T_{A},\Gamma _{MN}\right] =0.$ Finally there is a local
bosonic/fermionic extended kappa (super)symmetry under {\it left
multiplication} of $g$ with infinitesimal coset elements $K$ of the form
(take OSp$\left( N|2M\right) $ as an example for the matrix notation) 
\begin{equation}
\delta g=Kg,\quad K=\left( 
\begin{array}{cc}
0 & \xi \left( \tau \right) \\ 
\tilde{\xi}\left( \tau \right) & 0
\end{array}
\right) ,\quad Str\left( \Gamma _{MN}K\right) =0,\quad Str\left(
T_{A}K\right) =0,
\end{equation}
provided $\delta A^{ij}$ is non-zero as specified below, and $\xi _{\tilde{%
\alpha}}^{a}\left( \tau \right) $ has the form 
\begin{equation}
\xi _{\tilde{\alpha}}^{a}\left( \tau \right) =X_{i}^{M}\left( \Gamma
_{M}\kappa ^{ia}\right) _{\tilde{\alpha}}\,\,,  \label{xi}
\end{equation}
where the local $\kappa _{\tilde{\alpha}}^{ia}\left( \tau \right) $ are
unrestricted local parameters. Under such a transformation we have 
\[
\delta \pounds =0+\frac{2}{s}L^{MN}Str\left( \left[ \Gamma _{MN}\,,K\right]
\partial _{\tau }gg^{-1}\right) -\frac{1}{2}\delta A^{ij}X_{i}\cdot X_{j}, 
\]
One must have $L^{MN}\left( \Gamma _{MN}\xi \right) $ proportional to $%
X_{i}\cdot X_{j}$ so that $\delta A^{ij}$ can be chosen to cancel the
contribution from the first term. Indeed, the general form in (\ref{xi}) has
this property 
\begin{equation}
L^{MN}\left( \Gamma _{MN}\xi \right) =\varepsilon
^{kj}X_{k}^{M}X_{j}^{N}X_{i}^{R}\left( \Gamma _{MN}\Gamma _{R}\kappa
^{i}\right) =2\varepsilon ^{kj}X_{k}^{M}\left( \Gamma _{M}\kappa ^{i}\right)
X_{j}\cdot X_{i}
\end{equation}
since the three gamma term in $\Gamma _{MN}\Gamma _{R}=\Gamma _{MNR}+\Gamma
_{M}\eta _{NR}-\Gamma _{N}\eta _{MR}$ drops out due to the fact that the
indices $i,j,k$ take only two possible values.

Let us now specialize to a few cases of interest and use the gauge
symmetries to cut down the degrees of freedom to those in $d$ dimensions
given in the beginning of the previous section.

\begin{itemize}
\item  We start with the purely bosonic particle. Using the SO$\left(
d,2\right) $ local symmetry we can choose $g\left( \tau \right) =1$ for all $%
\tau .$ The Lagrangian (\ref{L}) is now expressed only in terms of $%
X^{M},P^{M}.$ We work in the basis $X^{M}=\left( X^{+^{\prime
}},X^{-^{\prime }},x^{\mu }\right) $ where $X^{\pm ^{\prime }}=\left(
X^{0^{\prime }}\pm X^{1^{\prime }}\right) /\sqrt{2}$ are lightcone type
coordinates for the extra two dimensions (similarly for $P^{M})$. Using the
Sp$\left( 2,R\right) $ local symmetry we can choose $X^{+^{\prime }}\left(
\tau \right) =1$ and $P^{+^{\prime }}\left( \tau \right) =0,$ and solve the
two constraints $X^{2}=0,$ $X\cdot P=0,$ to give $X^{-^{\prime }}=x^{2}/2$
and $P^{-^{\prime }}=x\cdot p.$ The Lagrangian reduces to the bosonic
particle Lagrangian in (\ref{L1}) without the fermions. When $g=1$ the
global SO$\left( d,2\right) $ that acts on the right of $g$ must be
compensated by a global transformation on the left of $g$ which also acts on 
$X_{i}^{M}.$ Thus the left/right transformations on $g$ get coupled and
become the 2T spacetime global SO$\left( d,2\right) $ Lorentz symmetry. When
the Sp$\left( 2,R\right) $ gauges are also fixed, this global symmetry acts
non-linearly on the remaining variables $\left( x^{\mu },p^{\mu }\right) $
and is then interpreted as the conformal symmetry of the massless particle.

\item  Next we discuss the superparticle in $d=3,4,6$ as in \cite{super2t}.
Using the local SO$\left( d,2\right) $ and local internal symmetries the
form of $g,$ for $G$ given in (\ref{G}), can be gauge fixed to $g\rightarrow
t=\exp \left( fermionic\,\ coset\right) $ parametrized by the spinor $\Theta
_{\tilde{\alpha}}^{a}\left( \tau \right) .$ This was the starting point in 
\cite{super2t}. Using the kappa gauge $\Gamma ^{+^{\prime }}\Theta =0,$ and
Sp$\left( 2,R\right) $ gauges $X^{+^{\prime }}=1,\,\ P^{+^{\prime }}=0$ it
was shown in \cite{super2t} that the 2T Lagrangian reduces precisely to the
superparticle Lagrangian in (\ref{L1}) for $d=3,4,6.$ The global symmetry $G$
becomes the non-linearly realized superconformal symmetry of the massless
superparticle in these special dimensions.

\item  For other supergroups the use of the local symmetries can never
reduce the degrees of freedom to only the superspace variables $\left(
x,p,\theta \right) .$ Generally there are more bosonic degrees of freedom.
It was speculated in \cite{super2t}\cite{liftM}\cite{toyM} that the extra
degrees of freedom can be associated with collective coordinates that
describe D-brane degrees of freedom in the particle limit. This is because
the superalgebra has central extensions with relations among the charges
such that BPS conditions are satisfied, as is the case for D-branes. This
point will be further elaborated in \cite{toyM} for the case of OSp$\left(
1|64\right) .$
\end{itemize}

\section{Twistors, supertwistors, oscillators}

To make the presentation as explicit as possible we will concentrate on the $%
d=3$ superparticle, rewritten in the $d+2=5$ two-time formalism. Hence we
will take $G=OSp\left( N|4\right) $ where Sp$\left( 4\right) =SO\left(
3,2\right) .$ However, we will begin more generally with $G=$OSp$\left(
M/2N\right) $ where Sp$\left( 2N\right) $ has an SO$\left( d,2\right) $
subgroup whose spinor representation has dimension $s=2N$. For example for $%
d=11$ in the toy M-model, $SO\left( 11,2\right) $ has a spinor with 64
components, and therefore we would consider OSp$\left( 1|64\right) .$

In the basis $X^{M}=\left( X^{+^{\prime }},X^{-^{\prime }},X^{\mu }\right) $%
, we define the following SO$\left( d,2\right) $ gamma matrices 
\begin{equation}
\Gamma ^{\pm ^{\prime }}=\pm \tau ^{\pm }\times 1,\quad \Gamma ^{\mu }=\tau
_{3}\times \gamma ^{\mu },\quad \left\{ \Gamma ^{M},\Gamma ^{N}\right\}
=2\eta ^{MN}.
\end{equation}
For the case of $d=3$ we take the following explicit form for the SO$\left(
2,1\right) $ gamma matrices 
\begin{equation}
\gamma ^{\mu }=\left( i\sigma _{2},\sigma _{1},\sigma _{3}\right)
\end{equation}

The group $G=$OSp$\left( M/2N\right) $ is characterized by $g\in G$ of the
form 
\begin{equation}
g^{-1}=\hat{C}g^{st}\hat{C}^{-1}
\end{equation}
where $g^{st}$ is the supertranspose of $g$ and $\hat{C}$ is the metric of
OSp$\left( M/2N\right) $. In supermatrix notation we can write 
\begin{equation}
\hat{C}=\left( 
\begin{array}{ll}
1_{M} & 0 \\ 
0 & C_{2N}
\end{array}
\right) ,\quad g=\left( 
\begin{array}{ll}
\alpha & \tilde{f}_{1} \\ 
f_{2} & A
\end{array}
\right) ,\quad g^{st}=\left( 
\begin{array}{cc}
\alpha ^{t} & f_{2}^{t} \\ 
-\tilde{f}_{1}^{t} & A^{t}
\end{array}
\right) ,\quad g^{-1}=\left( 
\begin{array}{cc}
\alpha ^{t} & \tilde{f}_{2} \\ 
f_{1} & \tilde{A}
\end{array}
\right)
\end{equation}
with the definitions 
\begin{equation}
\tilde{f}=f^{t}C_{2N}^{-1},\quad \tilde{A}=C_{2N}A^{t}C_{2N}^{-1}.
\end{equation}
Here $C_{2N}$ is the antisymmetric metric for Sp$\left( 2N\right) $, with
properties 
\begin{equation}
C_{2N}^{t}=-C_{2N}=C_{2N}^{-1},\quad \left( C_{2N}\right) ^{2}=-1_{2N}.
\end{equation}
Note the extra minus sign in $-\tilde{f}_{1}^{t}$ in the definition of
supertranspose $g^{st}$. This is necessary so that the supertranspose
operation has the property $\left( g_{1}g_{2}\right)
^{st}=g_{2}^{st}g_{1}^{st}.$ The necessity of the extra minus sign in $-%
\tilde{f}^{t}$ can be traced to the extra minus sign that arises from
anticommuting two fermions under transposition. Thus the parameters are
constrained by the relation 
\begin{eqnarray}
1 &=&g^{-1}g=\left( 
\begin{array}{cc}
\alpha ^{t} & \tilde{f}_{2} \\ 
f_{1} & \tilde{A}
\end{array}
\right) \left( 
\begin{array}{cc}
\alpha & \tilde{f}_{1} \\ 
f_{2} & A
\end{array}
\right) \\
\alpha ^{t}\alpha +\tilde{f}_{2}f_{2} &=&1_{N},\quad \alpha ^{t}\tilde{f}%
_{1}+\tilde{f}_{2}A=0,\quad f_{1}\alpha +\tilde{A}f_{2}=0,\quad f_{1}\tilde{f%
}_{1}+\tilde{A}A=1.
\end{eqnarray}
For infinitesimal group parameters the solution of these constraints are 
\begin{equation}
\delta \alpha ^{t}=-\delta \alpha ,\quad \delta f_{1}=-\delta f_{2}\equiv
\delta f,\quad \delta \tilde{A}=-\delta A
\end{equation}
Thus, one may write 
\begin{equation}
g=\exp \left( 
\begin{array}{cc}
\delta \alpha & -\delta \tilde{f} \\ 
\delta f & \delta A
\end{array}
\right)
\end{equation}
where $\delta f$ is any $M\times \left( 2N\right) $ fermionic matrix, $%
\delta \alpha $ is any $M\times M$ antisymmetric matrix and $\delta A$ is
any $\left( 2N\right) \times \left( 2N\right) $ symplectic matrix that
satisfies $C_{2N}\delta A^{t}C_{2N}^{-1}=-\delta A$.

\subsection{The global current}

Using Noether's theorem one finds that the global OSp$\left( N|2M\right) $
current of our model is 
\begin{equation}
J=g^{-1}Lg.
\end{equation}
where $L=\frac{1}{2}L_{MN}\Gamma ^{MN}$. At the classical level the current
satisfies 
\begin{equation}
J^{2}=g^{-1}Lgg^{-1}Lg=g^{-1}L^{2}g\sim 0.
\end{equation}
The vanishing is because of the Sp$\left( 2,R\right) $ constraints $%
X^{2}=P^{2}=X\cdot P=0$ at the classical level. As discussed elsewhere \cite
{conf} the treatment of constraints at the quantum level modifies this
result such that the Casimir operators Str$\left( J^{n}\right) $ computed
from the global currents are all non-zero but fixed at definite constants
that define a specific representation.

In the supermatrix notation given above the current is given by 
\begin{equation}
J=\left( 
\begin{array}{cc}
\alpha ^{t} & \tilde{f}_{2} \\ 
f_{1} & \tilde{A}
\end{array}
\right) \left( 
\begin{array}{ll}
0 & 0 \\ 
0 & L
\end{array}
\right) \left( 
\begin{array}{cc}
\alpha & \tilde{f}_{1} \\ 
f_{2} & A
\end{array}
\right) =\left( 
\begin{array}{ll}
\tilde{f}_{2}Lf_{2} & \tilde{f}_{2}LA \\ 
\tilde{A}Lf_{2} & \tilde{A}LA
\end{array}
\right)
\end{equation}
The parameters $\alpha $ have dropped out because of the internal gauge
symmetry. We will now choose further gauges that completely eliminate all
degrees of freedom from $X^{M},P^{M},$ shifting the dynamics to the degrees
of freedom in $g\left( \tau \right) .$ We will show that the remaining
degrees of freedom are supertwistors. Using the local SO$\left( d,2\right) $
and Sp$\left( 2,R\right) $ gauge symmetries we can map the two vectors $%
X^{M}\left( \tau \right) ,P^{M}\left( \tau \right) $ to the constants $%
X^{+^{\prime }}=1,$ $P^{-}=1$ with all other components zero $X^{M}=\delta
_{+^{\prime }}^{M},$ $P^{M}=\delta _{-}^{M}.$ These satisfy the constraints $%
X^{2}=P^{2}=X\cdot P=0.$ In this gauge we have 
\begin{equation}
L=\frac{1}{2}\Gamma ^{MN}L_{MN}=\Gamma ^{-^{\prime }}\Gamma ^{+}=\tau
^{-}\times \gamma ^{+}=\left( 
\begin{array}{ll}
0 & 0 \\ 
\gamma ^{+} & 0
\end{array}
\right) ,
\end{equation}
Plugging this form into the Lagrangian we find that the remaining degrees of
freedom are described by the following dynamics 
\begin{equation}
\pounds \sim Tr\left( \tilde{A}\tau ^{-}\gamma ^{+}\partial _{\tau }A+\tilde{%
f}_{2}\tau ^{-}\gamma ^{+}\partial _{\tau }f_{2}\right) .
\end{equation}
The current given above satisfies the commutation rules of OSp$\left(
N|2M\right) $ if we take the basic commutation rules 
\begin{equation}
\left\{ f_{\alpha }^{i},f_{\beta }^{j}\right\} =i\delta ^{ij}\hat{L}_{\alpha
\beta },\quad \left[ A_{\alpha }^{\,\,\gamma },A_{\beta }^{\,\,\delta }%
\right] =i\hat{L}_{\alpha \beta }\left( C^{-1}\right) ^{\gamma \delta }
\end{equation}
where $\hat{L}$ is defined by the relation $L\hat{L}L=L,$ and is given by 
\begin{equation}
\hat{L}=\tau ^{+}\times \gamma ^{-}.
\end{equation}
In the basis of gamma matrices we have chosen, the charge conjugation matrix
is $C_{4}=\tau _{1}\times C_{2}$. Using this $C_{4}$ we also parametrize the
4-column $f_{2}$ and the 4$\times $4 matrix $A$ in terms of 2 dimensional
blocks 
\begin{equation}
f_{2}^{i}=\left( 
\begin{array}{c}
\xi ^{i} \\ 
\chi ^{i}
\end{array}
\right) ,\quad A=\left( 
\begin{array}{ll}
a & b \\ 
c & d
\end{array}
\right) ,\quad \tilde{f}_{2i}=\left( 
\begin{array}{ll}
\tilde{\chi}_{i} & \tilde{\xi}_{i}
\end{array}
\right) ,\quad \tilde{A}=\left( 
\begin{array}{ll}
\tilde{d} & \tilde{b} \\ 
\tilde{c} & \tilde{a}
\end{array}
\right)
\end{equation}
where $\tilde{\xi}_{i}=\xi _{i}^{T}C_{2}^{-1}$ and $\tilde{a}%
=C_{2}a^{T}C_{2}^{-1}$, etc. We may now compute the Lagrangian and the
global OSp$\left( 8|4\right) $ currents more explicitly 
\begin{eqnarray}
\pounds &\sim &\frac{1}{2}\tilde{\xi}^{i}\gamma ^{+}\partial _{\tau }\xi
^{j}+\frac{1}{2}Tr\left( \tilde{b}\gamma ^{+}\partial _{\tau }a+\tilde{a}%
\gamma ^{+}\partial _{\tau }b\right) \\
J &=&\left( 
\begin{array}{lll}
\tilde{\xi}^{i}\gamma ^{+}\xi ^{j} & \tilde{\xi}^{i}\gamma ^{+}a & \tilde{\xi%
}^{i}\gamma ^{+}b \\ 
\tilde{b}\gamma ^{+}\xi ^{j} & \tilde{b}\gamma ^{+}a & \tilde{b}\gamma ^{+}b
\\ 
\tilde{a}\gamma ^{+}\xi ^{j} & \tilde{a}\gamma ^{+}a & \tilde{a}\gamma ^{+}b
\end{array}
\right) =\left( 
\begin{array}{ccc}
I^{ij} & -\tilde{Q}^{i} & -\tilde{S}^{i} \\ 
S^{j} & J & K \\ 
Q^{i} & P & -\tilde{J}
\end{array}
\right)
\end{eqnarray}
The canonical pairs are easily determined from the Lagrangian. The currents
are quadratic in the canonical pairs. They are reminiscent of the oscillator
formalism for supergroups \cite{barsgunaydin} and indeed we will make this
connection much clearer in the following paragraphs. The interpretation of
the various components of the currents are as follows. The 8$\times 8$ block 
$I^{ij}=\tilde{\xi}^{i}\gamma ^{+}\xi ^{j}$ corresponds to the generators of
SO$\left( 8\right) ,$ The 2$\times 8$ blocks $Q^{i}=\tilde{a}\gamma ^{+}\xi
^{i}$ and $S^{i}=\tilde{b}\gamma ^{+}\xi ^{i}$ represent 8 supercharges and
8 superconformal charges respectively. The 4$\times 4$ block including $%
J,P,K $ represents Sp$\left( 4\right) =SO\left( 3,2\right) $, with $J,P,K$
corresponding to Lorentz transformations and dilatations, translations and
conformal transformations respectively.

Note that the parameters $\chi ,c,d$ have dropped out from the Lagrangian
and the physical global currents. This is because of the extended local
kappa supersymmetry and bosonic gauge symmetries. The remaining parameters $%
\xi ,a,b$ are constrained by $gg^{-1}=1$, or $f_{2}\tilde{f}_{2}+A\tilde{A}%
=1 $ or 
\begin{equation}
\xi ^{i}\tilde{\xi}^{i}+a\tilde{b}+b\tilde{a}=0.
\end{equation}
This also guarantees $J^{2}=0.$ Furthermore, because 
\begin{equation}
\gamma ^{+}=\left( 
\begin{array}{cc}
0 & 1 \\ 
0 & 0
\end{array}
\right)
\end{equation}
is a projection operator, the currents depend only on a few of the
parameters in the 2$\times 2$ matrices $a,b$ and column matrix $\xi .$ This
too is because of the extended gauge symmetry. To further simplify the
expressions we define the entries in each matrix

\[
\xi ^{i}=\left( 
\begin{array}{c}
\lambda ^{i} \\ 
\theta ^{i}
\end{array}
\right) ,\quad a=\left( 
\begin{array}{cc}
a_{3} & a_{4} \\ 
a_{1} & a_{2}
\end{array}
\right) ,\quad b=\left( 
\begin{array}{cc}
b_{3} & b_{4} \\ 
b_{1} & b_{2}
\end{array}
\right) , 
\]
We find that the currents are given by the unrestricted parameters $\left(
\theta ^{i},a_{1},a_{2},b_{1},b_{2}\right) $ corresponding to the second row
of the matrices above, while the remaining parameters drop out from the
Lagrangian and currents. We will write $a_{\alpha }=\left(
a_{1},a_{2}\right) ,$ $a^{\alpha }=\left( a_{2},-a_{1}\right) ,$
raising/lowering indices $\alpha $ by using the Levi-Civita symbol $%
\varepsilon ^{\alpha \beta }$ which is the charge conjugation matrix. In
this notation the Lagrangian and currents take the form 
\begin{equation}
\pounds \sim \frac{1}{2}\theta ^{i}\dot{\theta}^{i}+b^{\alpha }\dot{a}%
_{\alpha },\quad J=\left( 
\begin{array}{lll}
\theta ^{i}\theta ^{j} & \theta ^{i}a_{\beta } & \theta ^{i}b_{\beta } \\ 
b^{\alpha }\theta ^{j} & b^{\alpha }a_{\beta } & b^{\alpha }b_{\beta } \\ 
a^{\alpha }\theta ^{j} & a^{\alpha }a_{\beta } & a^{\alpha }b_{\beta }
\end{array}
\right) =\left( 
\begin{array}{ccc}
I^{ij} & -\tilde{Q}^{i} & -\tilde{S}^{i} \\ 
S^{j} & J & K \\ 
Q^{i} & P & -\tilde{J}
\end{array}
\right)
\end{equation}
The basic non-zero commutation rules 
\begin{equation}
\left\{ \theta ^{i},\theta ^{j}\right\} =i\delta ^{ij},\quad \left[
a_{\alpha },b^{\beta }\right] =i\delta _{\alpha }^{\beta }
\end{equation}
determine the algebra of the conserved charges $J$. It is easily seen that
the charges form the OSp$\left( 8|4\right) $ superalgebra. In fact the form
of this construction is identical to the oscillator construction of
superalgebras \cite{barsgunaydin}. In the present case the real fermions $%
\theta ^{i}$ are SO$\left( 8\right) $ spinors and the canonical conjugates $%
a_{\alpha }$, or $b^{\alpha }$ are real Sp$\left( 2\right) $ spinors, where
Sp$\left( 2\right) =SO\left( 2,1\right) $ is the Lorentz subgroup of the
conformal group Sp$\left( 4\right) =SO\left( 3,2\right) $. In the usual
oscillator construction of \cite{barsgunaydin} one chooses the maximal
compact subgroup of OSp$\left( 8|4\right) ,$ which is SU$\left( 4\right)
\times SU\left( 2\right) \times \left( U\left( 1\right) \right) ^{2},$ and
takes a complex quartet of fermionic oscillators in the fundamental
representation of SU$\left( 4\right) \times U\left( 1\right) $ and a complex
doublet of bosonic oscillators in the compact subgroup SU$\left( 2\right)
\times U\left( 1\right) ,$ plus their hermitian conjugates. The relation
between these two is just a change of basis such that the Sp$\left( 4\right) 
$ quartet given in the Sp$\left( 2\right) $ basis $\left(
a_{1},a_{2},b_{1},b_{2}\right) $ is rexpressed in terms of the oscillators $%
A_{\alpha }=\left( a_{\alpha }+ib^{\alpha }\right) /\sqrt{2}$ as a Sp$\left(
4\right) $ quartet in an SU$\left( 2\right) \times U\left( 1\right) $ basis $%
\left( A_{1},\,A_{2},\,A_{2}^{\dagger },-A_{1}^{\dagger }\right) .$
Similarly the real SO$\left( 8\right) $ spinor $\theta ^{i}$ is re-expressed
in the complex SU$\left( 4\right) \times U\left( 1\right) $ basis in terms
of fermionic oscillators. This construction proves the unitarity of the
representation, and identifies the super conformal particle with the super
doubleton representation of OSp$\left( 8|4\right) .$ It is evident that the
same general arguments hold for OSp$\left( N|4\right) .$

Returning to the Sp$\left( 2\right) =SO\left( 3,2\right) $ spinors $%
a_{\alpha }$ or $b^{\alpha }$, we can interpret them as the twistor
representation of the particle canonical coordinates $x^{\mu },p^{\mu }$
\`{a} la Penrose as follows. To avoid complications due to quantum ordering,
we will discuss only the classical version of this interpretation. Also, we
concentrate only on the purely bosonic Sp$\left( 4\right) $ to avoid
complications with the supergroup. Starting with the generators $P_{\beta
}^{\alpha }=a^{\alpha }a_{\beta }$ which form a traceless 2$\times 2$
matrix, we identify the momentum as 
\begin{equation}
a^{\alpha }a_{\beta }=\left( \gamma ^{\mu }\right) _{\beta }^{\alpha
}\,p_{\mu }\,,
\end{equation}
Evidently it describes a massless particle, since $p^{2}\delta _{\beta
}^{\alpha }=a^{\alpha }a_{\gamma }a^{\gamma }a_{\beta }=0.$ The coordinate $%
x^{\mu }$ is defined by the following relation between the two spinors $%
a_{\alpha },b_{\beta }$ 
\begin{equation}
b_{\alpha }=x^{\mu }\left( \gamma _{\mu }\right) _{\alpha }^{\beta }a_{\beta
}.
\end{equation}
To show that this is indeed the case, consider the conformal generator $%
K^{\mu }$ written in the form $K_{\beta }^{\alpha }=K^{\mu }\left( \gamma
^{\mu }\right) _{\beta }^{\alpha }=b^{\alpha }b_{\beta }$ and insert the
expression for $b_{\alpha }$ 
\begin{eqnarray}
K^{\mu } &=&\frac{1}{2}b^{\alpha }\left( \gamma ^{\mu }\right) _{\alpha
}^{\beta }b_{\beta }=-\frac{1}{2}x^{\lambda }a^{\alpha }\left( \gamma
_{\lambda }\gamma ^{\mu }\gamma _{\nu }\right) _{\alpha }^{\beta }a_{\beta
}x^{\nu } \\
&=&-\frac{1}{2}x^{\lambda }p_{\sigma }Tr\left( \gamma ^{\sigma }\gamma
_{\lambda }\gamma ^{\mu }\gamma _{\nu }\right) _{\alpha }^{\beta }x^{\nu }=%
\frac{1}{2}x^{2}p_{\mu }-x\cdot px^{\mu }.
\end{eqnarray}
This result is the well known expression for the conformal generator
(avoiding quantum ordering, or fermionic contributions). Similarly we
compute the dimension operator $D=\frac{1}{2}b^{\alpha }a_{\alpha }$ and the
Lorentz generator $J^{\mu \nu }=b^{\alpha }\left( \gamma ^{\mu \nu }\right)
_{\alpha }^{\beta }a_{\beta }$%
\begin{eqnarray}
D &=&\frac{1}{2}b^{\alpha }a_{\alpha }=\frac{1}{2}x^{\mu }a^{\alpha }\left(
\gamma _{\mu }\right) _{\alpha }^{\beta }a_{\beta }=\frac{1}{2}x^{\mu
}p^{\nu }Tr\left( \gamma _{\mu }\gamma _{\nu }\right) =x\cdot p \\
J^{\mu \nu } &=&b^{\alpha }\left( \gamma ^{\mu \nu }\right) _{\alpha
}^{\beta }a_{\beta }=x^{\lambda }a^{\alpha }\left( \gamma _{\lambda }\gamma
^{\mu \nu }\right) _{\alpha }^{\beta }a_{\beta }=x^{\lambda }p_{\sigma
}Tr\left( \gamma _{\lambda }\gamma ^{\mu \nu }\gamma _{\sigma }\right)
=x_{\mu }p_{\nu }-x_{\nu }p_{\mu }
\end{eqnarray}
These are the correct expressions for the massless bosonic particle.

This makes it evident that the basis we have defined in terms of the super
variables $\left( \theta ^{i},a_{\alpha },b_{\alpha }\right) $ is indeed the
twistor basis for the conformal superparticle. In the presence of fermions
the relation between the spinors $b_{\alpha },a_{\alpha }$ is considerably
more complicated such that the correct generators $K^{\mu },J^{\mu \nu }$
emerge, as given in \cite{super2t}. The generalized relation including the
fermions is obtained by applying the kappa and other gauge transformations
that take the model from the fixed gauge $g\left( \tau \right) =t\left( \tau
\right) $ used in \cite{super2t} to the fixed gauge of the twistor formalism
described in this paper.

\section{Discussion and generalizations}

From the previous section, it is evident that the approach is applied in a
straightforward manner in $d=4,6$ using the supergroups $SU\left(
2,2|N\right) $ and OSp$\left( 8^{\ast }|N\right) $ respectively. The main
point is the shifting of the degrees of freedom from $X,P$ to $g$ or
vice-versa by using the gauge symmetries Sp$\left( 2,R\right) $ and SO$%
\left( d,2\right) .$ When $X,P$ are eliminated, the remaining gauge
symmetries reduce further the degrees of freedom in $g$ to the super twistor
variables for those dimensions. In particular, for $d=4$ we find agreement
with \cite{kallosh}. Similarly, twistors can be gauge transformed to
coordinate representation to describe the same system. Twistors were
invented as a means of describing conformal systems covariantly in linear
realizations. We have now shown that they are related to another linear
realization, namely Lorentz transformations in 2T physics.

In this paper we have mainly concentrated on a specific Sp$\left( 2,R\right) 
$ gauge choice, namely $X^{+^{\prime }}=1,$ $P^{+^{\prime }}=0$ which
relates to the SO$\left( d-1,1\right) $ covariant massless relativistic
particle. This is a particular embedding of $d$ dimensions inside $d+2$
dimensions, and corresponds to a particular 1T physics interpretation of the
2T theory. As we know from \cite{conf}-\cite{toyM} there are many other 1T
embeddings with different 1T physics interpretations. Such other Sp$\left(
2,R\right) $ gauges may now be combined with the present techniques of
shifting particle variables to twistor-like variables embedded in $g,$ and
thus find new twistor-like realizations of 1T dynamical systems, as well as
establish duality-type relations among them.

The reader will notice that the spacetime and internal subgroups of OSp$%
\left( 8|4\right) ,$ SU$\left( 2,2|4\right) ,$ OSp$\left( 8^{\ast }|4\right) 
$ were treated in an asymmetric manner in the coupling introduced in (\ref{L}%
). Since these supergroups describe the supersymmetries in AdS$_{7}\times
S_{4},$ AdS$_{5}\times S_{5},$ AdS$_{4}\times S_{7},$ respectively, one may
wonder if there is a more symmetric treatment of the AdS$\times $S spaces
that would apply to these cases. In fact, in addition to the spacetime $%
X^{M},P^{M}$ phase space we may introduce internal $Y^{I},K^{I}$ phase
space, define the internal angular momentum $L^{IJ}=Y^{I}K^{J}-Y^{J}K^{I}$
and couple it to the supergroup Cartan connection in the same manner as (\ref
{L}). When both $L^{MN}$ and $L^{IJ}$ have non-zero coupling it is possible
to maintain a kappa-type local supersymmetry as well as the Sp$\left(
2,R\right) $ local symmetry coupled to all the coordinates $\left(
X^{M},Y^{I}\right) $ and momenta $\left( P^{M},K^{I}\right) .$ This
variation leads to more interesting and intricate 2T models. In fact it is
even possible to couple to any a subgroup of the internal group SO$\left(
N\right) ,$ SU$\left( N\right) ,$ Sp$\left( N\right) $ that appear in (\ref
{G}). If only a subgroup $H$ is gauged, then only the corresponding degrees
of freedom can be removed. The remaining coset plays the role of harmonic
superspace recently discussed in \cite{ferrara}.

As mentioned earlier, a motivation for developing these techniques was the
study of the quantum system for the toy M-model that will be discussed
elsewhere \cite{toyM}. Having shown that the approach works and establishes
connections among previously better understood systems, it may now be used
to explore new systems.

\end{document}